\documentclass[runningheads,a4paper]{llncs}

\usepackage{amssymb}
\usepackage{graphicx}
\usepackage{verbatim} 

\usepackage{url}
\urldef{\mailsa}\path|michaelp@di.uoa.gr|

\newcommand{\USQL}{\textsf{USQL}}
\newcommand{\Proteus}{\textsf{Proteus}}

\begin{document}
\mainmatter  

\title{Development of a Language and its Enacting Engine for the Unified Discovery of Heterogeneous Services}
\titlerunning{Unified Discovery of Heterogeneous Services}

\author{Michael Pantazoglou\thanks{Dissertation Advisor: Aphrodite Tsalgatidou, Assist. Professor}}
\authorrunning{Michael Pantazoglou}

\institute{National and Kapodistrian University of Athens\\Department of Informatics and Telecommunications\\
\mailsa\\}
\maketitle

% Abstract
\begin{abstract}
Service orientation fosters a high-level model for distributed applications development, which is based on the discovery, composition and reuse of existing software services. However, the heterogeneity among current service-oriented technologies renders the important task of service discovery tedious and ineffective. This dissertation proposes a new approach to address this challenge. Specifically, it contributes a framework supporting the \emph{unified} discovery of heterogeneous services, with a focus on web, peer-to-peer, and grid services. The framework comprises a service query language and its enacting service discovery engine. Overall, the proposed solution is characterized by generality and flexibility, which are ensured by appropriate abstractions, extension points, and their supporting mechanisms. The viability, performance, and effectiveness of the proposed framework are demonstrated by experimental measurements. 
%\keywords{Service discovery, heterogeneous services, interoperability}
\end{abstract}

% DISSERTATION SUMMARY
\section{Dissertation Summary}
\label{section_dissertation_summary}

% Service-Oriented Computing and Service Discovery
\subsection{Service-Oriented Computing and Service Discovery}
\label{subsection_service_oriented_computing_service_discovery}
Thanks to its high level of abstraction and the promise of interoperability among heterogeneous systems, service-oriented computing \cite{HuhnsSingh2005} appeals to designers and developers of distributed software applications. According to the principles of this paradigm, an application is not necessarily built from scratch, but is rather composed by reuse of existing software services, which can be discovered and appropriately orchestrated to deliver the desired functionality. The service-oriented software engineering process involves interactions among three principal actors, as follows: 

\begin{itemize}
\item \emph{Service providers} expose their software solutions as services, and render them discoverable by publishing their descriptions to appropriate brokers
\item \emph{Service brokers} provide the appropriate infrastructure to facilitate publication and discovery of available services
\item \emph{Service consumers} first discover the services they need by searching the contents of one or more service brokers, and subsequently invoke them by interacting with the corresponding service providers, through well-defined interfaces
\end{itemize}

Clearly, the task of service discovery is essential to service-oriented computing, as it allows developers to find existing services that meet their application requirements, and reuse them accordingly, instead of implementing their functionality themselves. Indeed, service discovery spans the development life cycle of a service-oriented solution and is iteratively applied to improve its functionality and performance: 

\begin{itemize}
\item At an early stage, it facilitates the specification and assessment of requirements by software analysts
\item At design-time, it assists software designers and developers in identifying the building blocks of their application, either through bottom-up or top-down analysis 
\item At run-time, it supports the on-demand substitution of malfunctioning services with functionally similar, yet better performing ones 
\end{itemize}

All in all, service discovery contributes to the adoption of service orientation, as it allows businesses, organizations, and other entities to dynamically find each other, in terms of the functionalities they offer, and subsequently interact through the services they provide, or need. 

% Problem Description
\subsection{Problem Description}
\label{subsection_problem_description}
To date, web services constitute the prevailing instantiation of the service-oriented computing paradigm; they are widely in use, whereas all fundamental aspects in regards to their description, discovery and invocation are governed by well-defined standards. In recent years, however, a shift of other distributed computing technologies such as peer-to-peer systems and Grids toward service orientation has been observed \cite{JXTA,OGSA,WSRF}, bringing them closer to each other, and to the web services technology. Peer-to-peer services and grid services are currently published, discovered and utilized in the development of full-fledged service-centric applications, in the same way web services are. Still, the combined usage of all these services, a requirement that is met today in many application domains, quickly becomes an intricated task. 

From the perspective of service discovery, such shortfall is due to the lack of appropriate means to address the following challenges:

\begin{itemize}
\item \emph{Heterogeneity in service descriptions}. Service requesters need assistance in shaping their queries independently from the various service models and description formats. 
\item \emph{Heterogeneity in service discovery mechanisms}. Service discovery queries should be fired independently from the underlying mechanisms provided by the infrastructure, where services are published. Moreover, low-level technical details should be transparent to the service requester. 
\item \emph{Multi-dimensional query formation and evaluation}. Service discovery queries should reflect real-world requirements by encapsulating search criteria towards multiple functional and qualitative properties of a service. Then, appropriate matchmaking models and mechanisms are required to support the evaluation of such queries.
\item \emph{Technological volatility}. In order to ensure their long-term viability, service discovery languages and tools need to be flexible enough, so as to easily evolve and adapt to advances in services technologies.
\end{itemize}

% Related Work
\subsection{Related Work}
\label{subsection_related_work}
Despite the considerable volume of research results in service discovery, the various proposed approaches only partially address the aforementioned challenges, and thus they cannot be applied to the discovery of heterogeneous services. In the following paragraphs, we give an overview of related work focusing on the proposed query languages and engines, and we assess their limitations. 

Most of the currently existing languages for service discovery fail to address the heterogeneity in service descriptions and service discovery mechanisms. In \cite{WSDA,JinEtAl2007}, the authors propose the use of XQuery to express queries over service descriptions that are constructed according to specific schemas. Similarly, in \cite{StrouliaWang2005,CrassoEtAl2008}, service queries can only be evaluated against WSDL documents. In an effort to address the need for multi-dimensional query formulation, the approach proposed in \cite{Ran2003} defines certain extensions to the UDDI specification, still queries are only compliant with registries of that type. Also, numerous approaches that have adopted existing semantic service description languages to express service queries, as for example in \cite{SrinivasanEtAlSWSWPC2005}, are naturally confined to limited sets of available services.

Many service discovery frameworks have also embraced semantic web technologies to improve precision and recall of the matchmaking process, as it is reported in \cite{EleniusIngmarsson2004,MWSDI,OWLS-IDE}. However, such approaches mainly focus on support for the evaluation of functional search criteria, they are constrained to specific types of service brokers, and generally lack flexibility. Other efforts have tackled the challenge of multi-dimensional query evaluation \cite{ASD,G-QoSM}, or the heterogeneity in service discovery mechanisms \cite{DIRE,PYRAMID-S}. Even though the proposed system architectures are characterized by flexibility, they exclusively support the discovery of web services and thus their solution is not applicable to other types of services. 

% Contributions
\subsection{Contributions}
\label{subsection_contributions}
With this dissertation, we argue that, a \emph{unified} approach can tackle all aforementioned challenges and  can thereby successfully remedy any deficiency in the course of discovering heterogeneous services. Motivated by the identified gap in languages and tools that could support such an approach, we back this argument by proposing a complete framework comprising a high-level service query language called \USQL{} (\textsf{U}nified \textsf{S}ervice \textsf{Q}uery \textsf{L}anguage), and a flexible service discovery engine called \Proteus{}. 

The main contribution of our unified approach lies in the idea that service discovery should be orthogonal with respect to the existing service technologies. To this end, the proposed framework:

\begin{itemize}
\item supports the construction and evaluation of service discovery queries, as well as the formulation of their responses, independently from the type of the requested service, the format of the service description, and the kind of broker infrastructure, where the service has been published. 
\item provides a rich syntax for queries that can capture an open set of functional and qualitative requirements towards the requested service, in the form of search criteria and filters.
\item can be easily extended with additional elements and functionality, thereby attaining flexibility and immunity to the volatility of service technologies.
\end{itemize}

Thanks to the inherent generality and flexibility of both \USQL{} and \Proteus{}, the contributed unified approach ensures interoperability with a wide spectrum of service technologies, and thus it can be applied to the discovery of heterogeneous services in a variety of environments and settings. Consequently, the proposed framework does not impose any restrictions on service providers, in regards to their existing service descriptions, or on service brokers, in regards to their currently supporting discovery mechanisms.

%\begin{figure}
%\c entering
%\includegraphics[scale=0.40]{../Dissertation/Figures/Chapter_03/USQLRequirementElements.eps}
%\caption{Hierarchy of service requirement elements in USQL}
%\label{figure_usql_requirement_elements}
%\end{figure}

% RESULTS AND DISCUSSION
\section{Results and Discussion}
\label{section_results_discussion}
This section briefly reports and discusses the primary results of this dissertation, namely the \USQL{} language and the \Proteus{} engine, focusing on their salient features.

% The Unified Service Query Language
\subsection{The Unified Service Query Language}
\label{subsection_usql}
The \emph{Unified Service Query Language}, abbreviated to \USQL{}, provides a high-level meta-model and its accompanying algebra to support the \emph{unified} discovery of heterogeneous services as follows:

\begin{itemize}
\item Queries are formulated independently from the type of the requested service and the discovery mechanism of the broker, where services are published
\item Queries can be evaluated over an open set of service descriptions adhering to diverse languages and schemas
\item Queries may encapsulate requirements and filters towards multiple functional and non-functional service properties
\item Query results are uniformly constructed and may contain entries that correspond to various types of services
\end{itemize}

The above features are attained by \USQL{} through its meta-model, which, thanks to its generic conceptual basis \cite{AthanasopoulosEtAlSCC2006,TsalgatidouEtAlIJWSR2008}, has been defined at a high level of abstraction, and is thus positioned orthogonally to the various service types, as well as to the heterogeneous service description and discovery technologies. 

At the core of \USQL{} lies a hierarchy of abstracted elements that are used to model weighted (i.e. prioritized) requirements towards a service. On top of these abstractions, \USQL{} defines a series of concretized elements, which can explicitly express requirements towards the service capability, the service interface (i.e. input/output messages and their constituent elements), the underlying resource and its properties, as well as constraints to an open set of qualitative service properties. Besides, the meta-model is open, allowing for seamless accommodation of additional types of service requirements. Notably, all supported requirements can be described by text and/or be semantically annotated with the use of ontology concepts. Such duality in requirements expression gives \USQL{} a significant advantage over many other service discovery languages, which are constrained to the use of either keywords-based text or semantic annotations. 

With \USQL{}, service discovery queries are formulated as \emph{USQLRequest} documents, which may blend the flavors of the aforementioned types of requirements along with an open set of search filters. The latter are treated as \emph{hard} constraints aiming at reducing the number of service entries that are retrieved by the service discovery engine upon query execution. For instance, in the case of queries targeting web services, peer-to-peer services, or grid services, typical examples of search filters might constrain the service provider, classification, the peer or peer group identifier, or the name of the virtual organization.

In addition to the expression of search criteria and filters, a \USQL{} query may convey information regarding its search target(s). This way, queries may be directed to one or more explicitly specified and potentially heterogeneous service brokers, which are uniformly described with the use of a generic element provided by the \USQL{} meta-model. Such feature is particularly important in closed environments, where service discovery needs to be conducted in a controlled manner, or in settings, where the potential search targets are more than one and thus they need to be determined. 

For completeness purposes, and in opposition to other approaches, \USQL{} conveniently defines the structure of service discovery results, which take the form of \emph{USQLResponse} documents. Each such document constitutes the \emph{answer} of the service discovery engine to a \USQL{} query, and contains entries corresponding to the matches that were found, i.e. the service operations that adequately satisfied the expressed search criteria and filters. The information conveyed by a matched service entry serves two purposes: 

\begin{itemize}
\item Firstly, it provides human-intended details of the corresponding service operation, such as its overall degree of match, its provider, name, description, and service type, which aim at facilitating the service selection process. Along those lines, included in the matched service entry may be also details regarding the service properties that were constrained by the query, such as the service capability, interface, underlying resource, or its QoS.
\item Secondly, appropriate details allowing the immediate invocation of the corresponding service are included. Such information is meant to be interpreted by appropriate software, such as service composition execution engines or other service clients, in order to access and call the particular service operation. As invocation details primarily depend on the service type, \USQL{} defines a simple, generic structure to facilitate their expression in a uniform manner.
\end{itemize}

\USQL{} is mathematically grounded by an algebra \cite{PantazoglouEtAlWISE2006}, which underpins the evaluation of \USQL{} queries. Among the main features of this mathematical model are its independency from the various service description schemas, and its flexibility in terms of the matchmaking algorithms that are employed to match values of various types (e.g. text, ontology concepts, numerical values, etc.). Similarly, the algebra relies on external mechanisms for the evaluation of technology-dependent requirements supported by the meta-model, such as the ones referring to QoS service properties and data types, while it can seamlessly accommodate the evaluation of new types of requirements. 

According to the \USQL{} algebra, a query containing weighted service requirements and an advertisement containing the corresponding properties of a service are perceived as two ordered sets, $Q$ and $A$, respectively, such that requirement $r_{i} \in Q$ refers to property $p_{i} \in A$. To match all constituent pairs $(r_{i}, p_{i})$, a number of functions that correspond to the requirement types supported by the \USQL{} meta-model are defined. These functions are then appropriately combined in the context of a \emph{match calculator}, which is used to quantify the overall degree of match between $Q$ and $A$, taking into account the different weights assigned to the requirement elements $r_{i} \in Q$. The match calculator is based on the following intuitive assumptions:

\begin{itemize}
\item User requirements towards the desired functional and/or non-functional service properties are usually prioritized. Such priorities may be either explicitly specified by means of appropriate weights, or be implied by the nature of each requirement. Hence, satisfying requirements of higher priority is more important than satisfying the ones with lower priority.
\item The more important a service requirement is regarded, the more its degree of satisfaction affects the overall degree of match of a candidate service. Consequently, services failing to meet the top-prioritized requirements are left off of the query results. 
\end{itemize}

The \USQL{} meta-model has been implemented through a declarative XML syntax, which enables the use of the language by both humans and software agents. Besides, the \Proteus{} engine, which we briefly present in the following section, supports the processing of XML-based \USQL{} queries by providing a reference implementation of the \USQL{} algebra. 

% The Proteus Engine
\subsection{The Proteus Engine}
\label{subsection_proteus}
\Proteus{} is an engine designed to support the unified discovery of heterogeneous services, with the use of \USQL{} as query language. In early specifications of the engine \cite{PantazoglouEtAlICSOC2006,PantazoglouEtAlICWS2006}, queries were dispatched to the various heterogeneous search targets, which was a time- and resource-consuming task. To address the inflicted performance challenges, which arise from the potentially large numbers and distribution of service brokers as well as from the large volumes of published services, the latest version of \Proteus{} adopts a traditional search engine-like architecture. Specifically, the process of query execution is clearly decoupled from the process of visiting the various service brokers, retrieving and processing their contents. Overall, the engine's functionality is organized in two independent subsystems: 

\begin{itemize}
\item The \emph{Crawler} subsystem embodies a set of components, which are necessary for: (1) periodically visiting heterogeneous, yet already known to the system service brokers, with the aim of harvesting their published service descriptions; (2) processing the heterogeneous service descriptions, so as to generate service advertisements which adhere to a common format; and (3) appropriately storing and indexing the generated service advertisements into the  internal repository of \Proteus{}.
\item The \emph{Query Processor} subsystem provides users with an interface for unified discovery of heterogeneous services. The subsystem accepts queries submitted by users in the form of USQLRequest documents, extracts the expressed filters to retrieve service advertisements from the internal repository of \Proteus{}, and further evaluates their content against the expressed search criteria according to the \USQL{} algebra, in order to calculate the overall degree of match. The results of that process are consolidated into a USQLResponse document, which is returned to the requester, as a response to their query.  
\end{itemize} 

Among the most essential features of the \Proteus{} architecture is its openness, thanks to which the engine, aligned with the flexibility of \USQL{}, effectively addresses the heterogeneity of existing service description and discovery technologies. At the core of the system lies a powerful extensibility mechanism, which allows for seamless accommodation of additional functionality in the form of \emph{plug-ins}. This mechanism comprises a registry, where meta-information on each contributed plug-in is maintained, and a selector, which is commonly used by both subsystems to find and dynamically instantiate pluggable components. The \Proteus{} engine can be seamlessly extended by plug-ins of various kinds, which are briefly described as follows: 

\begin{itemize}
\item \emph{Harvester} plug-ins abstract the Crawler subsystem from the heterogeneity of existing service discovery mechanisms, and are employed to retrieve the contents of various service brokers. Selection of an appropriate harvester is based on the type of the corresponding broker, which the Crawler needs to visit. Each harvester plug-in accepts as input an object containing information about the broker to be visited, such as its access details, and returns as output a list of objects that correspond to the service publications found in that broker.
\item \emph{Parser} plug-ins provide the Crawler with the ability to parse the contents of an open set of heterogeneous service description documents. For each such document, the appropriate parser is identified by the namespace of the schema used for its formulation. (e.g. WSDL \cite{WSDL}, OWL-S \cite{OWL-S}, etc.). Each parser plug-in accepts as input an object that corresponds to a service publication, and returns a list of service advertisements abiding by a common, \USQL{}-compliant format, which result from the parsing process.
\item \emph{QoS Matcher} plug-ins address the diversity of QoS requirements that can be set in a \USQL{} query, and are employed by the Query Processor to support their evaluation upon matchmaking. The Query Processor identifies and selects the appropriate QoS matcher to be used for each QoS requirement according to its denoted name (e.g. Availability, Reliability, ResponseTime, etc.). All these components implement a common interface accepting as input two objects that correspond to the requirement and the respective service property, and returning as output their degree of match.  
\item \emph{Datatype Matcher} plug-ins are used by the Query Processor to calculate the degree of match among requested and advertised data types. They are specifically involved in the matchmaking of requirements towards (1) the input/output messages of a service, where the requester may have specified the desired data type for each constituent message element; and (2) the resource of a grid service, where the desired data types of the constituent resource properties may be specified. Each datatype matcher encapsulates all details related to the supported data type namespace (e.g. XSD \cite{XMLSchema}, JSON \cite{JSON}, etc.), which it corresponds to, thereby rendering \Proteus{} independent from those technologies. 
\end{itemize}

In addition to the flexibility acquired through the above mentioned extensibility mechanism, \Proteus{} is also characterized by modularity rendering the internal components of the engine independent from each other. Thus, changes to a particular component do not affect the remainder of the subsystem, which it belongs to. Such design is particularly useful in the case of the components responsible for the matchmaking of textual values and ontology concepts: in accordance to the \USQL{} algebra, the details of their respective implementing algorithms are isolated from the main matcher component, thereby allowing for their seamless substitution if needed. 

% Evaluation Results
\subsection{Evaluation Results}
\label{subsection_evaluation_results}
Our approach to service discovery was evaluated through a number of studies and experiments. Starting out with \USQL{}, we assessed its generality and suitability in the discovery of heterogeneous services by conducting a detailed survey on existing standards. Firstly, we compared \USQL{} to the interfaces provided by prevailing service discovery mechanisms. Secondly, we mapped the \USQL{} meta-model to the schemes of widely used service description languages. The goal of such evaluation was to demonstrate that a unified approach in the formulation of service discovery queries is feasible with minimum defects in terms of expressiveness and compatibility with the various existing service-oriented technologies. As the survey results indicated, \USQL{} covers all types of search filters and criteria that are currently supported by the existing service discovery mechanisms. Also, a high degree of compatibility with the most prominent service description languages was established, verifying that \USQL{} is indeed applicable to the discovery of a wide range of services. 

The effectiveness of our approach in service matchmaking, through the \USQL{} algebra, was quantified by experimenting on the precision and recall of the search results corresponding to queries for different kinds of web, peer-to-peer, and grid services. In order to estimate the contribution of each distinct type of requirements supported by \USQL{}, all queries that were used in the experiment were incrementally constructed. The experimental results showed that, on the one hand, the performance of our approach in terms of precision and recall is satisfactory, although it also depends on the performance of the various external matchmaking algorithms that are embodied in the \USQL{} algebra for the evaluation of text values, ontology concepts, data types, and QoS requirements. On the other hand, it became evident that, the ability of \USQL{} to combine different kinds of requirements in a single query generally enhances the accuracy of the search results. 

The evaluation of our approach was concluded with a number of experiments on the performance of \Proteus{}. According to the measurements retrieved through the use of a prototype implementation of the engine, both subsystems perform in a satisfactory manner. Further, through replication of the Crawler and the Query Processor, we verified that, despite heterogeneity, \Proteus{} scales well as the numbers of search targets, published services, and incoming queries increase. Overall, the abstraction and flexibility required for the unified discovery of heterogeneous services did not inflict the performance of our system. 

\section{Conclusions}
\label{section_conclusions}

We proposed a unified approach to the discovery of heterogeneous services, which is realized by a framework containing a query language called \USQL{} and its enacting service discovery engine called \Proteus{}. Both the language and the engine are characterized by a high level of abstraction and flexibility, thereby being orthogonally positioned to the heterogeneous service-oriented technologies. Initially implemented as part of the SODIUM platform \cite{SODIUM}, our framework was successfully applied to the unified discovery of web services, grid services, and peer-to-peer services, which were subsequently composed into applications in the domains of healthcare and crisis management \cite{TsalgatidouEtAl2006}. The extensibility of \USQL{} was also exercised in a preliminary effort to support the expression of requirements towards the behavior of composite services, as reported in \cite{PantazoglouEtAl2007}. It is also worth mentioning that, besides its reference implementation in \Proteus{}, \USQL{} was also adopted by two other proposed frameworks for semantically enhanced web service publication and discovery, namely PYRAMID-S \cite{PYRAMID-S} and DIRE \cite{DIRE}.

The ideas revolving around the query language and its enacting service discovery engine presented in this dissertation can be extended in several research directions. Some indicative examples are given below:

\begin{itemize}
\item First of all, it would be interesting to assess the applicability of our approach in discovering additional types of services that have become popular in the last few years, such as RESTful services \cite{Vinoski2008}, or OSGi services \cite{PreuveneersBerbers2008}.
\item Further, assuming the availability of appropriate standards, the \USQL{} meta-model could be extended in the future to accommodate information related to the \emph{context} of the requested service. 
\item Finally, given the proliferation of Web 2.0 and social networking technologies, the exploitation of techniques such as \emph{tagging} could enhance the results of service discovery, through the collective intelligence that is gradually shaped by service requesters in regards to the services semantics. Some preliminary results towards that direction have already been reported in \cite{PantazoglouTsalgatidou2008}. 
\end{itemize}

It is in our belief that, the accomplished flexibility of both \USQL{} and \Proteus{} will facilitate the undertaking of such endeavors in the future.

%\subsection{Replication Mechanism}
%\label{subsection_replication_mechanism}

% Evaluation Results
%\section{Evaluation}
%\label{section_evaluation_results}

% Discussion{Discussion}
%\section{Discussion}
%\label{section_discussion}

\bibliographystyle{splncs}
\bibliography{DissertationSummary}

\end{document}